\documentclass[aps,prd,10pt,twocolumn,nofootinbib,groupedaddress,superscriptaddress,amsfonts,floatfix]{revtex4-1}

\usepackage[colorlinks=true,urlcolor=blue,linkcolor=,citecolor=blue]{hyperref}
\usepackage{amsmath,amssymb,amstext,amsbsy,amsfonts,amsthm,graphicx,microtype,dsfont}
\usepackage[capitalize]{cleveref}

\def\nn{\nonumber}
\def\({\left(}
\def\){\right)}
\def\[{\left[}
\def\]{\right]}

\def\d{{\rm d}}

\usepackage{xcolor}
\definecolor{darkgreen}{cmyk}{0.85,0.2,1.00,0.2}

\begin{document}

\pagestyle{plain}

\title{Gravitational waves from gauge preheating}

\author{Peter Adshead}
\email{adshead@illinois.edu}
\affiliation{Department of Physics, University of Illinois at Urbana-Champaign, Urbana, Illinois 61801, USA}

\author{John T. Giblin, Jr.}
\email{giblinj@kenyon.edu}
\affiliation{Department of Physics, Kenyon College, Gambier, Ohio 43022, USA}
\affiliation{Department of Physics, Case Western Reserve University, Cleveland, Ohio 44106, USA}

\author{Zachary J. Weiner}
\email{zweiner2@illinois.edu}
\affiliation{Department of Physics, University of Illinois at Urbana-Champaign, Urbana, Illinois 61801, USA}

\begin{abstract}
We study gravitational wave production during Abelian gauge-field preheating following inflation.
We consider both scalar and pseudoscalar inflaton models coupled directly to Abelian gauge fields via either a dilatonic coupling to the gauge-field kinetic term or an axial coupling to a Chern-Simons term.
In both cases gravitational waves are produced efficiently during the preheating phase, with a signature louder than most cosmological signals.  These gravitational waves can contribute to the radiation energy budget of Universe at a level which will be probed by upcoming cosmic microwave background experiments through $N_{\rm eff}$. For axially coupled fields the resulting gravitational wave spectrum is helically polarized---a unique feature that can be used to differentiate it from other stochastic gravitational wave backgrounds.
We compute the gravitational topological charge and demonstrate that gauge preheating following axion inflation may be responsible for the matter-antimatter asymmetry of the Universe via gravitational leptogenesis. 
\end{abstract}

\maketitle

\section{Introduction}
\label{sec:intro}

Inflation~\cite{Guth:1980zm, Linde:1981mu, Albrecht:1982wi, Linde:1983gd} is the leading paradigm for the very early Universe.
As well as solving the flatness and horizon problems, inflation provides a consistent framework for the generation of a nearly scale-invariant spectrum of density fluctuations~\cite{Hawking:1982cz, Guth:1982ec, Bardeen:1983qw}.
At the end of inflation, the Universe must transition from its super-cooled, post-inflationary state to one filled with matter and radiation in order to begin the hot big bang phase---that is, the Universe must be reheated.

The phase of reheating can be extremely violent, beginning with a phase of preheating where the coherent oscillation of the background inflaton field excites parametric resonances in both itself and in fields to which it is coupled~\cite{Traschen:1990sw, Shtanov:1994ce, Kofman:1994rk,Kofman:1997yn}.
This period of rapid particle production is highly inhomogeneous and generically generates gravitational waves with large energy densities~\cite{Khlebnikov:1997di, Easther:2006gt, Easther:2006vd, Easther:2007vj, GarciaBellido:2007dg, Dufaux:2007pt, Dufaux:2010cf, Bethke:2013aba, Figueroa:2013vif, Bethke:2013vca, Figueroa:2016ojl, Figueroa:2017vfa}.
While preheating into scalars is well-established, comparatively less work has been done on direct preheating into gauge fields~\cite{GarciaBellido:1999sv, Rajantie:2000fd, Smit:2002yg, GarciaBellido:2003wd, Skullerud:2003ki, Tranberg:2003gi, DiazGil:2007dy, DiazGil:2008tf, GarciaBellido:2008ab, Dufaux:2010cf, Deskins:2013lfx, Adshead:2015pva, Lozanov:2016pac, Adshead:2016iae, Adshead:2017xll}. Generically, gauge preheating is more efficient and violent than scalar preheating, and we anticipate large associated gravitational-wave energy densities.

Gravitational-wave production from gauge fields at preheating has been studied previously in slightly different contexts.
In particular, the work of Ref.~\cite{Dufaux:2010cf} studied gravitational wave production from gauge fields and cosmic strings during preheating after hybrid inflation.
That work specifically focused on an Abelian-Higgs model where the symmetry breaking scalar field (the Higgs) was charged under a local U(1) symmetry.
The production of Nielsen-Olesen vortices and associated cosmic strings play an important role in the subsequent generation of gravitational radiation and imprint multiple scales into the spectrum.
Further, the work in Ref.~\cite{Figueroa:2014aya, Figueroa:2016ojl} considered gravitational-wave production from the generation of gauge-field modes from the decay of the standard model Higgs into the standard model fields after inflation.
In those scenarios, the Higgs contributes only a subdominant component of the total energy density, and the resulting spectrum is very small.
Finally, Ref.~\cite{Tranberg:2017lrx} considered a tachyonic transition of an SU(2)-Higgs like system at the electroweak scale.
These studies differ from the work here in the nature of the gauge-field interactions. 

It is well known that axial couplings from axion-driven (natural) inflation~\cite{Freese:1990rb,Adams:1992bn} to gauge fields leads to the exponential production of helically polarized gauge bosons during the inflationary phase~\cite{Garretson:1992vt, Prokopec:2001nc, Anber:2009ua,Barnaby:2010vf,Adshead:2013qp,Cheng:2015oqa}.
These helical gauge bosons can rescatter off the inflaton condensate leading to interesting phenomenology such as the production of non-Gaussian density perturbations~\cite{Barnaby:2011vw, Barnaby:2011qe}, primordial black holes~\cite{Linde:2012bt, Bugaev:2013fya, Cheng:2015oqa}, magnetic fields~\cite{Garretson:1992vt, Adshead:2016iae}, and (chiral) gravitational waves~\cite{Sorbo:2011rz, Barnaby:2011qe, Cook:2013xea,Adshead:2013qp} (for a review, see~\cite{Pajer:2013fsa}). These effects are exponential in the axion velocity and can become large near the end of inflation and during the reheating phase that follows. 

In this work we extend the studies of Refs.~\cite{ Deskins:2013lfx, Adshead:2015pva, Lozanov:2016pac, Adshead:2016iae} to compute the spectrum of gravitational waves produced during gauge preheating.
We find that gravitational waves are efficiently produced during reheating in both the axial and dilatonic scenarios. In the case of the axial coupling we find that the resulting spectrum of stochastic gravitational waves is helically polarized, which is potentially interesting for future stochastic gravitational wave observatories~\cite{Smith:2016jqs}.
Further, we demonstrate that these chiral gravitational waves lead to an appreciable topological charge of the right order of magnitude to explain the matter-antimatter asymmetry of the Universe via gravitational leptogenesis~\cite{Adshead:2017znw}.
In both the axial and dilatonic cases, the energy density in the resulting gravitational waves saturates the bounds of Ref.~\cite{Giblin:2014gra}, and are potentially large enough to be probed by future cosmic microwave background (CMB) experiments, such as CMB-S4~\cite{Abazajian:2016yjj}.

We work in natural units where $\hbar = c =1$; however, we retain the Planck mass, $m_{\rm pl} = 1.22 \times 10^{19} \, {\rm GeV}$.

\section{Background and Conventions}\label{background}

In this work, we consider models of inflation driven by a single scalar or pseudoscalar field (the inflaton).
The inflaton is coupled to a U(1) gauge field\footnote{Throughout we refer to ``electric'' and ``magnetic'' fields; however, the gauge fields we consider here need not correspond to that of Maxwell's electromagnetism.} via either a dilatonic-like coupling $W(\phi)$ to the field-strength tensor, or an axial coupling $X(\phi)$ to the Chern-Simons term, with the action\footnote{Greek letters here and throughout denote four dimensional Lorentz indices and Roman letters from the middle of the alphabet are used to denote spatial indices. Repeated lower spatial indices are summed using the Kronecker delta.}
\begin{align}\label{eqn:action}
	S = \int \d^4 x \sqrt{-g}\Bigg[&\frac{m_{\rm pl}^2}{16 \pi}R - \frac{1}{2}\partial_\mu\phi\partial^\mu \phi - V(\phi)\\ \nn&-\frac{W(\phi)}{4}F_{\mu\nu}F^{\mu\nu} - \frac{X(\phi)}{4} F_{\mu\nu}\tilde{F}^{\mu\nu}\Bigg].
\end{align}
We work with the background, homogeneous Friedmann-Lema\^itre-Robertson-Walker (FLRW) universe in conformal time with mostly-plus conventions, $\d s^2 = -a^2(\d\tau^2 - \d {\bf x}^2).$
Our Fourier convention is
\begin{align}\label{fourierConvention}
	f(\textbf{k})
	= \int \d^3x \, f({\bf x}) e^{i{\bf k}\cdot{\bf x}}.
\end{align}
The field strength tensor $F_{\mu\nu}$ and its dual $\tilde{F}^{\mu\nu}$ are given by their standard expressions
\begin{align}
	F_{\mu\nu} = \partial_\mu A_\nu - \partial_{\nu}A_{\mu}\; \;\text{ and } \;\;\tilde{F}^{\mu\nu} = \frac{1}{2}\epsilon^{\mu\nu\alpha\beta}F_{\alpha\beta},
\end{align}
where $\epsilon^{\mu\nu\alpha\beta}$ is the completely antisymmetric tensor and our convention is $\epsilon^{0123} = 1/\sqrt{-g}$.
We consider dilatonic and axial couplings of the form
\begin{align}
	W(\phi) &= \exp\left(- \frac{\beta}{M} \phi \right), \quad
	X(\phi) = \frac{\alpha}{M} \phi,
\end{align}
where $M$ is a free parameter with dimensions of mass, and $\alpha$ and $\beta$ are order unity and dimensionless.
In practice we study the two cases separately.
That is, we take either the combination $X=0$ and $W = \exp(- \beta \phi / M)$, or $W=1$ and $X = \alpha\phi/M$.

For definiteness, we consider the potential $V(\phi)$ for the simplest type of chaotic inflation, $V(\phi) = \frac{1}{2}m^2\phi^2$~\cite{Linde:1981mu}.\footnote{Although this model is now disfavored by data at the 95\% confidence level, we do not expect that our results are sensitive to this choice.}
The amplitude of the scalar spectrum fixes the scale to be $m \approx 1.06 \times 10^{-6}\,m_{\rm pl}$~\cite{Planck:2013jfk}.

\subsection{Equations for the background spacetime}

The equations of motion for the background metric (i.e., the scale factor) are the usual Friedmann equations 
\begin{align}
	\label{friedmann1}
	\mathcal{H}^2 \equiv \(\frac{a'}{a}\)^2 &= \frac{8\pi}{3m_{\rm pl}^2}\rho a^2, \\
	\label{friedmann2}
	 \frac{a''}{a}
	 = \mathcal{H}'-\mathcal{H}^2 &= -\frac{4\pi}{m_{\rm pl}^2}(\rho+p) a^2,
\end{align}
where here and throughout a prime (${}^\prime$) denote a derivative with respect to conformal time.
The (homogeneous) energy density and pressure of the universe are
\begin{align}\label{eqn:rhoandp}
	\rho &\equiv \rho(\tau) = \langle \rho_\phi \rangle+\langle \rho_{\rm A} \rangle, \\
	p &\equiv p(\tau) = \langle p_\phi \rangle+ \langle p_{\rm A}\rangle,
\end{align}
where the scalar components are
\begin{align}\label{eqn:scalarrho}
	\rho_\phi = & \frac{1}{2}\frac{\phi'{}^2}{a^2} +\frac{1}{2}\frac{(\partial_i \phi)^2 }{a^2} + V(\phi), \\
	p_\phi = & \frac{1}{2}\frac{\phi'{}^2}{a^2} - \frac{1}{6}\frac{(\partial_i \phi)^2 }{a^2} - V(\phi)
\end{align}
and the gauge-field components are
\begin{align}
	\rho_{A} = & \frac{W(\phi)}{2 a^4}(\partial_0 A_i - \partial_i A_0)^2 + \frac{W(\phi)}{4 a^4}(\partial_i A_j - \partial_j A_i)^2
\end{align}
and $p_{A} = \rho_A/3$.
Note that the axial coupling produces no stress-energy since it is topological.
The angle brackets $\langle \cdots \rangle$ in \cref{eqn:rhoandp} denote an average over the spatial hypersurface,
\begin{align}
	\langle \rho({\bf x},\tau) \rangle = \frac{1}{V}\int_{V} \d^3 x \, \rho({\bf x},\tau),
\end{align}
where 
\begin{align}
	V = \int_V \d^3 x
\end{align}
is the (comoving) volume of the hypersurface.
For our simulation $V = L^3$, where $L$ is the length of the box.

\subsection{Field equations}

The equation of motion for the inflaton is the Klein-Gordon equation with additional source terms from its interactions with the gauge fields,
\begin{align}
	\Box \phi + a^2\frac{d V}{d \phi} & = - \frac{a^2}{4} \frac{d X}{d \phi} F_{\mu\nu} \tilde{F}^{\mu\nu} - \frac{a^2}{4} \frac{d W}{d \phi} F_{\mu\nu} F^{\mu\nu},
\end{align}
where $\Box \equiv \partial^2/\partial\tau^2+2\mathcal{H}\partial/\partial\tau- \nabla^2$ is the wave operator in FLRW spacetime.
The equations of motion for the gauge field follow from the variation of the action, reading
\begin{align}
	\partial_{\mu}\left( \sqrt{-g}W(\phi)F^{\mu\nu}+\sqrt{-g}X(\phi)\tilde{F}^{\mu\nu} \right) = 0.
\end{align}
We work in the Lorenz gauge, $\partial^\mu A_{\mu} = 0$, where the variation with respect to $A_0$ produces the equation of motion
\begin{align}\label{A0EOM}
	W(\phi) \left( \partial_0^2 A_0 - \partial_j \partial_j A_0 \right)
	= \partial_k \left( W(\phi) F_{k 0} + X(\phi) \tilde{F}_{0 k} \right).
\end{align}
The Lorenz gauge choice serves as a physical constraint on the system of gauge fields and fixes the gauge up to a solution of the homogeneous wave equation.
Using this gauge choice to replace $A_0''$ with $\partial_i A_i'$ in \cref{A0EOM}, we obtain Gauss's law,
\begin{align}\label{gaussLaw}
	W(\phi) \left( \partial_i A_i' - \partial_j \partial_j A_0 \right)
	= \partial_k \left( W(\phi) F_{k 0} + X(\phi) \tilde{F}_{0 k} \right),
\end{align}
which additionally constrains the initial gauge field velocities.

The temporal gauge is the typical choice for simulations implementing lattice gauge theory; in this gauge, Gauss-violating modes are unstable~\cite{Moore:1996wn,Moore:1996qs}.
However, Gauss's law is a symmetry of the lattice-gauge formulation, meaning that if the initial conditions satisfy Gauss's law (at or near machine precision), then the evolution scheme exactly preserves that satisfaction.
In this work we evolve the continuum gauge potentials themselves; working in Lorenz gauge, we observe that constraint violation remains bounded and does not spoil the dynamics of our simulations.

\section{Gravitational Wave Production}

The violent production of matter fields during the final stages of inflation and the early stages of reheating potentially leads to the generation of large metric fluctuations, including the Newtonian potentials.
However, for the purposes of this work, we only track the evolution of the transverse-traceless (gravitational wave) parts of the metric and leave the study of the gravitational potentials for future work.

We write the perturbed line element (about the FLRW background) as
\begin{align}
	\d s^2 = -a^2(\tau) \left[\d\tau^2 -\left(\delta_{ij} + h_{ij}\right)\d x^i \d x^j\right],
\end{align}
where $\partial_i h_{ij} = h_{ii} = 0$ is a transverse traceless perturbation of the spatial metric, and we have set the scalar and vector perturbations to zero.
The Einstein equation gives the equation of motion for $h_{ij}$
\begin{align}\label{eqn:GWinhomo}
	{h^{\prime \prime}_{ij}}- \nabla^2 h_{ij}+ 2\mathcal{H}{h^{\prime}_{ij}} = 16\pi G \,S_{ij}^{\rm TT}.
\end{align}
In this expression, $S_{ij}^{\rm TT}$ is the transverse-traceless projection of the anisotropic stress tensor, obtained via
\begin{align}
	S_{ij}^{\rm TT} = \left(P_{il}P_{jm} - \frac{1}{2}P_{ij}P_{lm}\right)T_{lm},
\end{align}
where the transverse-traceless projector is
\begin{align}\label{projector}
	P_{ij}=\delta_{ij} - \frac{k_i k_j}{k^2}.
\end{align}
The stress-energy tensor has two contributions:\footnote{We ignore the contribution of the gravitational waves themselves to the transverse-traceless part of the stress tensor.} one from the scalar-field sector,
\begin{align}\label{fullTphi}
	T_{ij}^\phi = \partial_i \phi \partial_j \phi - a^2 \delta_{ij}\left[\frac{1}{2}\partial_\mu\phi \partial^\mu \phi + V(\phi)\right],
\end{align}
and one from the gauge-field sector,
\begin{align}\label{fullTa}
	T_{ij}^A = W(\phi) \left[ F_{i\alpha} F_{j\beta} g^{\alpha\beta} - \frac{g_{ij}}{4} \left( F_{\mu\nu}F^{\mu\nu} \right) \right].
\end{align}
In terms of the usual gauge-invariant electric and magnetic fields, the gauge-field stress tensor is
\begin{align}\label{reducedTa}
	T_{ij}^A = - \frac{ W(\phi) }{a^2} \left[ E_i E_j + B_i B_j + \frac{\delta_{ij}}{2} \left( \textbf{E}^2 + \textbf{B}^2 \right) \right].
\end{align}
Since the trace of $T_{ij}$ does not contribute anisotropic stress, it is sufficient in practice to only calculate the parts of $T_{ij}^\phi$ and $T_{ij}^A$ which are not explicitly traceless.
This amounts to dropping the parts of \cref{fullTphi,reducedTa} that are proportional to the Kronecker $\delta$.
From there, we apply \cref{projector} to obtain the fully transverse and traceless tensor $S_{ij}^{TT}$.

\subsubsection{Gravitational wave polarization}

The gravitational waves can be expanded into Fourier modes as
\begin{align}
	h_{ij}({\bf x}, \tau) =\!\sum_{\lambda = \pm} \int \frac{\d^3 k}{(2\pi)^3} \Pi^{*}_{ij, \lambda} ({\bf k})h^{\lambda}_{ k}(\tau)e^{-i {\bf k}\cdot{\bf x}}+{\rm c.c},
\end{align}
where we work with the circular polarization tensors
\begin{align}\label{GW-pol-definition}
	 \Pi^{*}_{ij, \pm} ({\bf k}) \equiv \epsilon^{(\pm)}_i ({\bf k})\epsilon^{(\pm)}_j ({\bf k}),
\end{align} 
where $\vec{\epsilon}\,{}^{(\pm)} ({\bf k})$ are the helicity vectors, which satisfy the relations ${\bf k}\cdot\vec{\epsilon}\,{}^{(\pm)} ({\bf k}) = 0$, $i{\bf k}\times \vec{\epsilon}\,{}^{(\pm)} ({\bf k}) = \pm k \vec{\epsilon}\,{}^{(\pm)} ({\bf k})$, $\vec{\epsilon}\,{}^{(\pm)} ({\bf k})\cdot \vec{\epsilon}\,{}^{(\pm)} ({\bf k}) = 0$, and $\vec{\epsilon}\,{}^{(\pm)} ({\bf k})\cdot \vec{\epsilon}\,{}^{(\mp)} ({\bf k}) = 1$.
Specifically, if $\hat{k} = (\sin\theta\cos\phi, \sin\theta\sin\phi, \cos\theta)$, then
\begin{align}\label{polarizationTensor}
	\vec{\epsilon}\,{}^{(\pm)} = \frac{1}{\sqrt{2}}(\cos\theta\cos\phi\mp i\sin\phi,\cos\theta \sin\phi\pm i\cos\phi, -\sin\theta).
\end{align}
We can then extract the gravitational wave polarization by projecting using the polarization tensors.

\subsubsection{Gravitational wave energy density}

The energy density of gravitational waves is~\cite{Misner:1974qy}
\begin{align}
	T_{\mu\nu}^{\rm gw} = \frac{m^2_{\rm pl}}{32\pi}\left\langle h_{ij,\mu} h_ {ij,\nu}\right\rangle,
\end{align}
where there is an implicit sum over $i$ and $j$.
The corresponding energy density is
\begin{align}
	\rho_{\rm gw} = \frac{T_{00}^{\rm gw}}{a^2} = \frac{m^2_{\rm pl}}{32\pi a^2}\left\langle h^\prime_{ij} {h^\prime_{ij}} \right\rangle.
\end{align}
Using Parseval's theorem,
\begin{align}
	\int \d^3k \,\left\vert f(\textbf{k}) \right\vert^2 = \left( 2 \pi \right)^3 \int \d^3x \, f({\bf x})^2,
\end{align}
we can rewrite the energy density in momentum space as
\begin{align}\nonumber
	\rho_{\rm gw}(k) &= \sum_{ij} \frac{m^2_{\rm pl}}{32\pi a^2} \frac{1}{L^3} \int \d^3x\, \left\vert h^\prime_{ij}({\bf x},\tau) \right\vert^2 \\ 
	&= \sum_{\lambda} \frac{m^2_{\rm pl}}{64 \pi^3 a^2} \frac{1 }{L^3} \int k^3\,\d \ln k \, \left\vert h_{k}^{\lambda}{}^\prime(\tau) \right\vert^2,
\end{align}
where $L$ is the length of the box.
We can use \cref{friedmann1} to write the fractional energy density in gravitational waves as
\begin{align}\label{GWdensitySpec}
	\Omega_\mathrm{gw}(k)
	\equiv \frac{1}{\rho} \frac{\d \rho_{\rm gw}}{\d \ln k}
	= \frac{1}{24\pi^2 L^3}\frac{k^3}{\mathcal{H}^2} \sum_{\lambda} \left\vert h^{\lambda}_k{}^\prime(\tau)) \right\vert^2.
\end{align}
Finally, the gravitational wave frequency that would be observed today is 
\begin{align}\label{frequencyToday}
	f \approx 6.0 \times 10^{10} \frac{k_{\rm phys}}{ \sqrt{m_\mathrm{pl} H}} \, \mathrm{Hz},
\end{align}
where $k_{\rm phys}$ is the {\sl physical} wave number and $H$ is the Hubble parameter evaluated at the time when the spectrum is being computed.
The transfer function to obtain today's gravitational wave amplitude is~\cite{Easther:2006vd}
\begin{align}
	\Omega_\mathrm{gw,0}(f) h^2
	&= \Omega_{\mathrm{gw},\mathrm{e}}(f) \left(\frac{g_0}{g_\ast}\right)^{1/3}\Omega_{\mathrm{r},0} h^2.
\end{align}
These formulas assume that the Universe has been radiation dominated since the time of emission until matter/radiation equality.

\section{Numerical scheme}\label{numerics}

We follow a substantial history of using numerical techniques to simulate classical field theories in discretized spacetimes.
The seminal numerical implementation, {\sc LatticeEasy}~\cite{Felder:2000hq}, was quickly followed by a family of codes, {\sc Defrost}~\cite{Frolov:2008hy}, {\sc PSpectRE}~\cite{Easther:2010qz}, {\sc HLattice}~\cite{Huang:2011gf}, and {\sc GABE}~\cite{Child:2013ria,Deskins:2013lfx}, among others.
In this work, we use \textsc{STELLA} (\textsc{pseudoSpecTral EvoLver on LAttices}), a new, GPU-accelerated code which builds off of the structure of \textsc{GABE}, and was used in a basic form in~\cite{Amin:2018xfe}.

As with most lattice methods, the fully nonlinear equations of motion governing the evolution of fields are evolved via the {\sl method of lines}. This scheme discretizes space onto a three-dimensional grid of length $L$ with $N$ points per dimension and numerically integrates all degrees of freedom at all $N^3$ sites.
Like \textsc{GABE}, \textsc{STELLA} uses a Runge-Kutta method for this time integration since our problem is not phase-separable as required by symplectic methods (as used by most of the aforementioned codes).
Here, however, we use a fourth-order Runge-Kutta scheme.

We evolve the scale factor using \cref{friedmann2}, with \cref{friedmann1} a constraint tracking the energy-conservation of the simulation.
For the homogeneous evolution, the energy and pressure are averaged over the grid at each stage of the integration.
Finally, we compute the generation of gravitational waves by solving the sourced, inhomogeneous (but linear) partial differential equation for the transverse-traceless part of the metric, \cref{eqn:GWinhomo}, in momentum space, similar to the implementation in~\cite{Easther:2007vj}.
We compute the source of \cref{eqn:GWinhomo} in configuration space, Fourier transform it, and project it onto the transverse-traceless space using \cref{projector}.

The main difference between the current software and prior methods is the computation of spatial derivatives.
Most (but not all) of the above cited packages implement standard finite-difference stencils to compute spatial derivatives; on the other hand, \textsc{STELLA} implements a spectral collocation method.
This routine computes spatial gradients and Laplacians in Fourier space by first Fourier transforming the fields, computing the $k$-space derivative, and inverse Fourier transforming the result back to position space.

The cost of performing so many discrete Fourier transforms is justified in several ways.
Spectral collocation methods are generally exponentially convergent, and therefore are the best possible approximation to spatial derivatives.
The Fourier basis is the correct choice for our periodic and regularly-spaced grid, allowing us to rely on optimized Fast Fourier Transform libraries for the majority of the computation.
In addition, pseudospectral gradients suffer none of errors incurred by finite-differencing that arise from computing the transverse-traceless projection via \cref{projector} (see~\cite{Huang:2011gf,Figueroa:2011ye} for details).

Lastly, \textsc{STELLA} is written in CUDA~\cite{CUDA} for implementation on NVIDIA GPUs, achieving runtimes one to two orders of magnitude shorter than multi-threaded CPU methods.
The implementation is currently limited to a single GPU, restricting accessible problem sizes to $N=256$; a future multi-GPU implementation will alleviate this restriction.

Our procedure for setting initial conditions follows Refs.~\cite{Deskins:2013lfx,Adshead:2015pva}: for a given initial power spectrum, random fluctuations are set in momentum space, drawing mode amplitudes from a Rayleigh distribution (via a Box-Muller transformation) and phases uniformly from $[0, 2\pi)$.
Having fully specified our initial conditions with these two variables, we set the fields' time derivatives to satisfy the source-free Klein-Gordon equation in a homogeneously expanding spacetime.
We detail the choice of initial power spectra for each model below.

Our initial conditions must satisfy both our chosen gauge condition and Gauss's law, \cref{gaussLaw}.
To satisfy the Lorenz gauge, we choose $A_0 = A_0' = 0$ and use \cref{projector} to remove the longitudinal component from $A_i$.
We satisfy Gauss's law on the initial slice with a relaxation method analogous to that used in~\cite{Adshead:2017xll}; we evolve the $A_i'$ through a fictitious time variable to dissipate violation of Gauss's law.

Finally, we use the window function
\begin{align}\label{window}
	F(k) = \frac{1}{2} \left( 1-\tanh(s(k-k_\ast)) \right)
\end{align}
to smoothly cut off (initial) modes larger than $k_\ast$.
The smoothness is parametrized by $s$ and the cutoff is set as a fraction of the Nyquist frequency of a lattice with $N$ points per side and length $L$,
\begin{align}
	k_\mathrm{Nyquist} = \frac{N \sqrt{3}}{2} \frac{2\pi}{L}.
\end{align}
This window function further improves the overall stability of the simulation by removing modes which are both the least well-resolved and the least physically relevant.
The ability to tune the smoothness of the window enables us to mitigate any possible issues from discontinuities in the initial power spectra.

In all of the simulations presented here, we use $N = 256$ grid points per side, a box length $L = 7.5$, and a fixed time step $dt = dx / 20$.
(Note that the horizon size at the end of inflation is typically $\mathcal{H} \sim 1-2 \, m^{-1}$).
We choose an initial cutoff $k_\ast = k_\mathrm{Nyquist} / 2$ with smoothing scale $s = L/8\pi$.
The scale factor is normalized such that $a = 1$ at the end of inflation, i.e.~when $\ddot{a} = 0$. The Friedmann constraint, \cref{friedmann1}, is satisfied to better than one part in $10^{6}$ over the course of the simulations.

\subsubsection{Dilatonic couplings}

For the dilatonic case, $W = W(\phi)$ and $X = 0$, we follow the strategies employed in~\cite{Deskins:2013lfx}.
The inflaton $\phi$ is initialized at the end of inflation with homogeneous value, $\left\langle \phi\right\rangle \approx 0.20 \,{m_{\rm pl}}$ and velocity, $\langle\dot{\phi}\rangle \approx -1.42 \times 10^{-7}\, m^2_{\rm pl}$, which is consistent with an inflaton mass $m = 10^{-6} \,m_{\rm pl}$.
The inflaton field is then seeded with a realization of the Bunch-Davies vacuum, so that the power spectrum of the field is
\begin{align}
	\left\langle \left\vert \phi_k \right\vert^2 \right\rangle = \frac{1}{2\omega_k},
\end{align}
where $\omega_k = \sqrt{k^2 + a^2 m^2}$.
For this model, the spatial components of the gauge fields are also initialized in the Bunch-Davies vacuum before implementing the strategies described above to satisfy the gauge constraints.
We simulate couplings $\log_{10} \beta = 1.70$, $1.74$, $1.78$, $1.82$, and $1.86$ (or $\beta \approx 50.1$, $56.0$, $60.2$, $66.1$, and $72.4$).

\subsubsection{Axial couplings}

In the case of an axial coupling to a Chern-Simons term, $W = 1$ and $X = X(\phi)$, we follow the procedure of~\cite{Adshead:2015pva, Adshead:2016iae}.
During axion inflation, axially-coupled gauge fields with wave numbers $k/aH < \xi$, where $\xi = \alpha \dot{\phi}/(2HM)$, are exponentially enhanced due to their interaction with the rolling homogeneous mode of the inflaton (see, for example,~\cite{Sorbo:2011rz}).
Therefore, many of the scalar- and gauge-field modes that are relevant for our simulation are not well-described as Bunch-Davies at the end of inflation.
Our strategy in this case is to numerically integrate the linearized equations of motion for the two gauge field polarizations, $A^\pm(k)$, from when each mode was deep inside the horizon (when it is still Bunch-Davies) until a certain number of $e$-folds before the end of inflation.
In this way, we capture the early stage of polarized amplification in the linear regime, but begin the lattice simulation early enough to ensure all relevant non-linear effects are captured.
As was chosen in~\cite{Adshead:2015pva,Adshead:2016iae}, we end the linear mode evolution and begin the lattice simulation 2 $e$-folds before the end of inflation.
We use the power spectra $\left\langle \vert A^\pm(k) \vert^2 \right\rangle$ obtained from the linear evolution for initial conditions, projecting with \cref{polarizationTensor} to obtain the components $A_i$.
The corresponding initial homogeneous field values for the inflaton are taken from the background evolution.
Finally, we simulate couplings $\alpha = 40$, $45$, $50$, $55$, $60$, and $65$.

\section{Results}

\begin{figure}[t!]
	\includegraphics[width=.99\columnwidth]{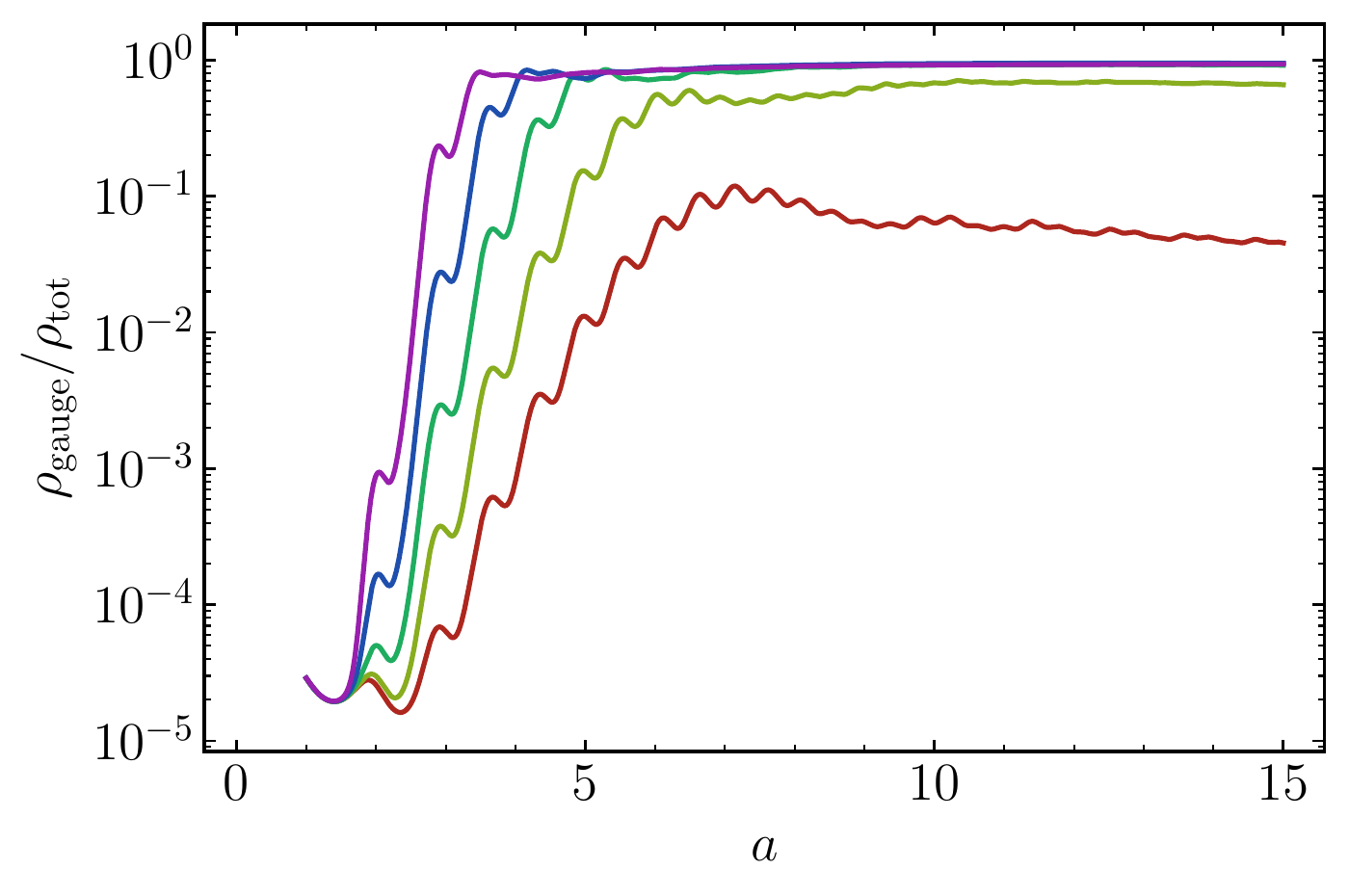}
	\includegraphics[width=.99\columnwidth]{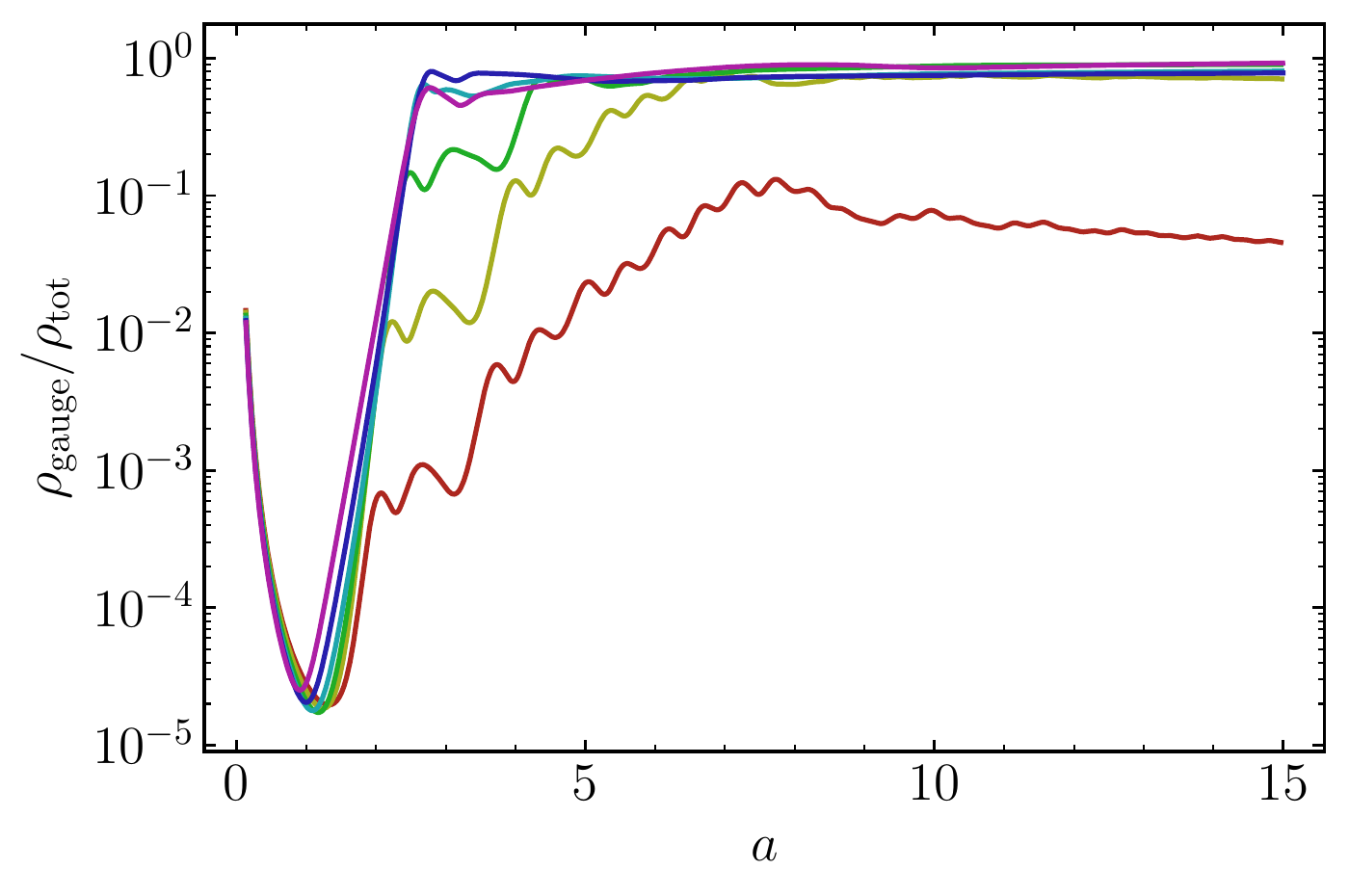}
	\caption{
		The ratio of energy density in the gauge fields to the total energy density, $\rho_\mathrm{gauge} / \rho_\mathrm{tot}$.
		The top panel displays the results for the dilatonic coupling, with colors red to purple denoting $\log_{10} \beta = 1.70$, $1.74$, $1.78$, $1.82$, and $1.86$. 
		The bottom panel depicts the results for the axial coupling, with red through purple corresponding to couplings $\alpha = 40$, $45$, $50$, $55$, $60$, and $65$.
		Note that in both plots $a=1$ corresponds to the end of inflation.
		}\label{fig:fracGaugeEnergy}
\end{figure}

We begin by showing the consistency of our numerical approach with previous results.
In \cref{fig:fracGaugeEnergy} we reproduce the results of~\cite{Deskins:2013lfx,Adshead:2015pva}, plotting the fraction of the total energy density residing in the gauge fields over the course of the simulation for both the dilatonic and axial cases.
We note that our results are insensitive to the initial realization of fluctuations and are consistent with lower-resolution ($N^3 = 128^3$) simulations.

\begin{figure}[t]
	\includegraphics[width=.99\columnwidth]{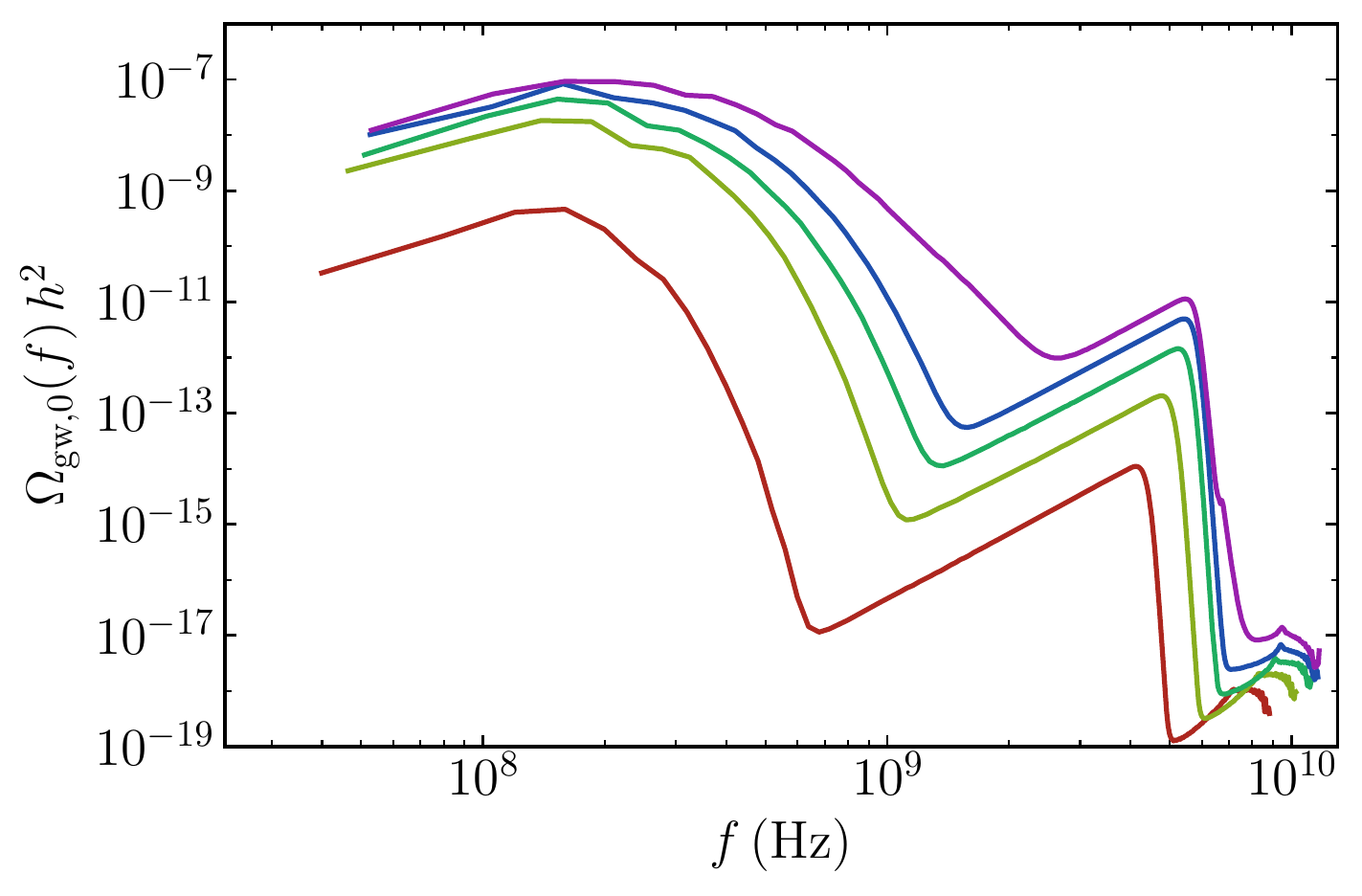}
	\caption{
		The energy density in gravitational waves today produced by at preheating by a dilatonically coupled gauge field, evaluated at $a = 15$.
		From red to purple $\log_{10} \beta = 1.70$, $1.74$, $1.78$, $1.82$, and $1.86$, or $\beta \approx 50.1$, $56.0$, $60.2$, $66.1$, and $72.4$.
		}\label{fig:dilatonGWPlot}
\end{figure}

\begin{figure}[t]
	\includegraphics[width=.99\columnwidth]{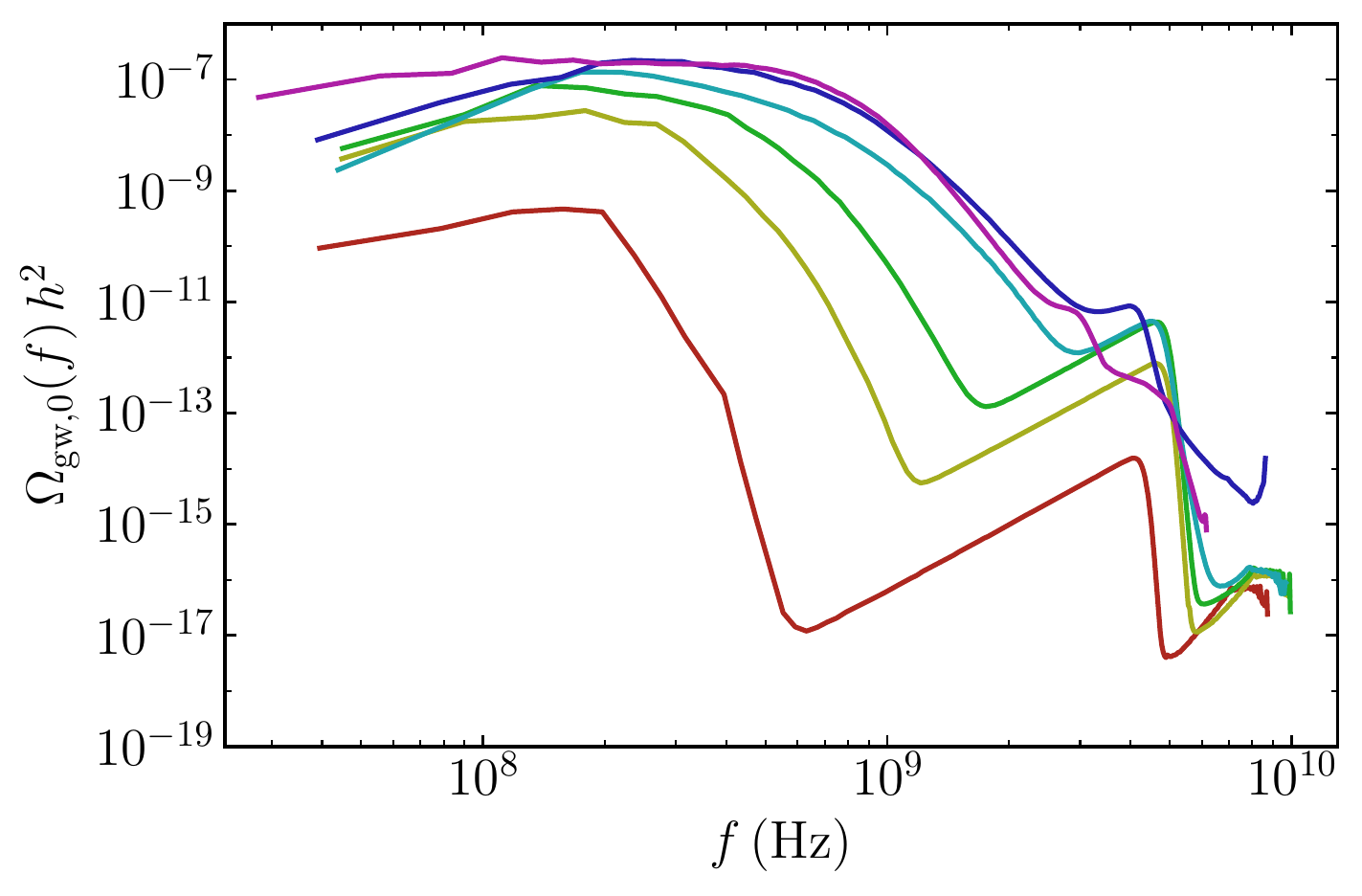}
	\caption{
		The energy density in gravitational waves today produced by at preheating by an axially coupled gauge field, evaluated at $a = 15$.
		From red to purple $\alpha = 40$, $45$, $50$, $55$, $60$, and $65$.
		}\label{fig:axionGWplot}
\end{figure}

\Cref{fig:dilatonGWPlot,fig:axionGWplot} display the resulting energy density today as a function of frequency today for gravitation waves produced by dilatonically- and axially-coupled gauge fields, respectively, during preheating.
These spectra are computed at the end of the simulation, $a=15$, although amplification ends well before this point.
Note that the broad features of the gravitational wave signal are remarkably similar in the two cases.
As the coupling is increased, the amplitude of the signal increases, the range of modes which are amplified broadens, and the peak moves to higher frequencies.
Increasing the couplings in both models amplifies gauge-field modes to an increased maximal $k$, since (to linear order) the coupling tunes the cutoff of the tachyonic resonance band.
These larger-momentum gauge bosons subsequently produce larger-momentum gravitational waves.
In the axial case (\cref{fig:axionGWplot}) we observe that $\alpha = 60$ and $65$ saturate the peak amplitude; additionally, in the simulation with $\alpha = 65$, reheating completes faster, resulting in a larger Hubble rate at the end of reheating which subsequently reduces the frequency that would be observed today (see \cref{frequencyToday}).

One notable feature in the results of both \cref{fig:dilatonGWPlot,fig:axionGWplot} is that the spectra at large wave numbers (beyond the physical peak) grow as $k^4$ (up to the point cut off by the window function applied to the initial conditions).
This is precisely the contribution of the unamplified vacuum modes to the spectrum, and contributes a UV divergence ($\sim \Lambda^4$, where $\Lambda$ is the UV cutoff) to the total energy in gravitational radiation.
In reality, this contribution to the stress-energy of the Universe is removed by renormalization and does not represent physical gravitational waves.

It is important to look at such a large signal with a degree of skepticism.
Most gravitational wave signals from cosmological processes saturate around $\Omega_{\mathrm{gw},0} h^2 \sim 10^{-10}$ which can be understood from `rule-of-thumb' calculations as in~\cite{Giblin:2014gra}.\footnote{See also~\cite{Amin:2014eta}.}
To that end, we need to understand why this signal may be so loud.
In the language of~\cite{Giblin:2014gra}, the gravitational wave signal today is
\begin{align}
\Omega_{\rm gw, 0}\approx 2.3\times 10^{-4} \alpha^2 \beta w^2 \frac{k_*}{\sigma} \left(\frac{H_p}{k_*}\right)^2,
\end{align}
where $\alpha$ is the fraction of the energy in the source compared to the total energy density of the Universe at the time of the process, $\beta$ measures how anisotropic the source is and $w$ is approximately the equation of state of the Universe at the time of the process.
In~\cite{Giblin:2014gra}, it was assumed that the source was Gaussian with a peak, $k_*$, and width, $\sigma$; both entered into the approximation as ratios that include the Hubble parameter at the time of the process, $H_p$.
The general advice in~\cite{Giblin:2014gra} chose fiducial parameters for these ratios and set estimates for optimistic, realistic, and pessimistic limits.
We cannot rely on these fiducial values here because the tachyonic processes at work in these models excite modes that are very close to horizon-sized, as can be seen in~\cite{ Adshead:2015pva}.
Using Fig.~1 of ~\cite{Adshead:2015pva} as a guide, we see that $k_*$ can be as small as $3$-$5\,m$ (in the axial case), which is only about a decade away from
\begin{align}
H_p \approx \sqrt{\frac{8\pi}{3 m_{\rm pl}^2}\left(\frac{1}{2}\dot{\phi}^2_0 + \frac{1}{2}m^2\phi_0^2\right)} \approx 0.5 \,m.
\end{align}
We can also approximate the width of the Gaussian to be about the same order of magnitude as the peak, $\sigma \sim 5\,{\rm m}$.
Taking $h = .68$ and the optimistic parameters, $\alpha = 1$, $\beta = 0.1$ and $w=1/3$, we get an optimistic estimate for the peak height to be
\begin{align}
\Omega_{\rm gw, 0}(f)\approx 10^{-8}.
\end{align}
where the estimate changes by a factor of a few when we go from $k_\ast=3\,m$ to $k_\ast=5\,m$.
Therefore we see that the enhancement of near--horizon-sized modes so quickly after inflation can create a gravitational wave signal a few orders of magnitude larger than we expect from a parametric instability.
These estimates are consistent with our most efficient results above.

Such a large signal puts the detection of gravitational waves from preheating within the reach of ground-based interferometers. In this work we choose the inflationary scale $m = 10^{-6} \, m_\mathrm{pl}$ to fit the amplitude of the scalar spectrum for chaotic inflation, for which reason the frequencies of the generated gravitational waves lie far from those to which LIGO is sensitive.
However, the amplitude of this signal will remain (relatively\footnote{The mass scale $m$ only enters the simulations via the initial amplitude of the Bunch-Davies spectrum; results should be fairly insensitive to this amplitude due to the preheating's dramatic amplification of modes.}) invariant when changing $m$, while the emitted frequencies are proportional to this scale~\cite{Easther:2006vd}.
For this reason, these preheating dynamics after low-scale inflation could in principle be detected by LIGO. Advanced LIGO's peak sensitivity is on the order of $\Omega_{\mathrm{gw},0}(f) h^2 \sim 10^{-10}$, which is several orders of magnitude \emph{lower} than that the amplitude produced by the simulations which achieve complete reheating.
aLIGO's peak sensitivity lies around $f \sim 50 \, \mathrm{Hz}$, which would probe inflationary scales $\sim 10^6 \, \mathrm{GeV}$.
Note that the subsequent expansion history of the Universe also affects the gravitational-wave transfer function; we assume the Universe is radiation dominated after emission until matter-radiation equality. Since preheating into gauge fields naturally leads into radiation domination after inflation, this approximation is well-justified.

Further, the total energy density in gravitational waves, i.e.,
\begin{align}
	\Omega_{\mathrm{gw},0} h^2
	= \int \d \ln k \, \frac{1}{\rho} \frac{ \d \rho_{\mathrm{gw},0} }{\d \ln k},
\end{align}
is constrained by CMB measurements.
If we assume that there are no light degrees of freedom beyond the standard model that contribute to the radiation density during the formation of the CMB, we can directly translate the constraint on $N_{\rm eff}$ onto a constraint on $\Omega_{\mathrm{gw},0}(f) h^2$ via~\cite{Maggiore:1999vm}
\begin{align}
	\frac{ \Omega_{\rm gw,0} h^2}{\Omega_{\gamma,0} h^2} &= \frac{7}{8} \left( \frac{4}{11} \right)^{4/3} \Delta N_{\rm eff},
\end{align}
where $\Omega_{\gamma,0}h^2 = 2.47 \times 10^{-5}$ is the present energy density in photons and $\Delta N_{\rm eff} =  (N_{\rm eff} - 3.046)$. Planck limits $\vert \Delta N_{\rm eff} \vert \lesssim 0.33$~\cite{Ade:2015xua}, which constrains the energy density to $\Omega_{\mathrm{gw},0} h^2 \lesssim 1.85 \times 10^{-6}$.
Next-generation CMB experiments, such as CMB-S4, will probe $\Delta N_{\rm eff}$ to a level of $\sigma( N_{\rm eff}) \sim 0.02-0.03$~\cite{Abazajian:2016yjj} and potentially constrain the gravitational wave energy density to
\begin{align}
	\Omega_{\mathrm{gw},0} h^2 \lesssim 1.12 - 1.68 \times 10^{-7}.
\end{align}
In \cref{tab:gw-integrated} we list the final value of $\Omega_{\mathrm{gw},0} h^2$ for each coupling; upcoming experiments could constrain the axion-gauge field coupling $\alpha < 55$ and the dilaton-gauge field coupling $\beta < 72.4$.
\begin{table}[t]
\begin{tabular}[t]{cc}	
	$\alpha$ & $\Omega_{\mathrm{gw},0} h^2$  \\
	\hline
	40 & $5.5 \times 10^{-10}$ \\
	45 & $3.4 \times 10^{-8}$ \\
	50 & $9.5 \times 10^{-8}$ \\
	55 & $1.6 \times 10^{-7}$ \\
	60 & $3.2 \times 10^{-7}$ \\
	65 & $5.4 \times 10^{-7}$
\end{tabular}
~
\begin{tabular}[t]{cc}
	$\beta$ & $\Omega_{\mathrm{gw},0} h^2$  \\
	\hline
	50.1 & $4.2 \times 10^{-10}$ \\
	56.0 & $2.0 \times 10^{-8}$ \\
	60.2 & $4.5 \times 10^{-8}$ \\
	66.1 & $8.5 \times 10^{-8}$ \\
	72.4 & $1.3 \times 10^{-7}$
\end{tabular}
\caption{The fraction of the total energy density of the simulation in gravitational waves, $\Omega_{\mathrm{gw},0} h^2$, for the axial coupling, $\alpha$, (left) and the dilatonic coupling, $\beta$ (right).}\label{tab:gw-integrated}
\end{table}
A more sophisticated forecast, such as that of~\cite{Pagano:2015hma}, obtains constraints as low as $\Omega_{\mathrm{gw},0} h^2 \lesssim 7.6 \times 10^{-8}$, which would probe $\alpha \approx 50$ and $\beta \approx 66.1$.

\subsection{Gravitational wave polarization}

During axion-driven inflation, the rolling axion preferentially amplifies one gauge-field polarization (in the linear regime).
These gauge fields in turn lead to the production of gravitational waves through their contribution to the anisotropic stress.
That is, scattering of helically-polarized gauge bosons produces gravitational waves~\cite{Sorbo:2011rz}. Because they are helically polarized and angular momentum is conserved in their scattering, these helical gauge bosons result in a chiral gravitational-wave spectrum---the amplitude of one helical polarization of the resulting gravitational-wave spectrum is larger than the other. 

During gauge preheating after axion inflation, the axion oscillates about the minima of its potential, amplifying each gauge-field helicity in turn.
Further, scattering of helical gauge bosons off of fluctuations in the axion background leads to the production of the other (subdominant) helicity~\cite{Adshead:2015pva}.
However, generically the final spectrum of gauge bosons is significantly polarized, and we therefore expect to produce a polarized spectrum of gravitational radiation during preheating.
In this section we study the polarization of the resulting gravitational wave spectra produced in our simulations.

\begin{figure}[t]
	\includegraphics[width=.99\columnwidth]{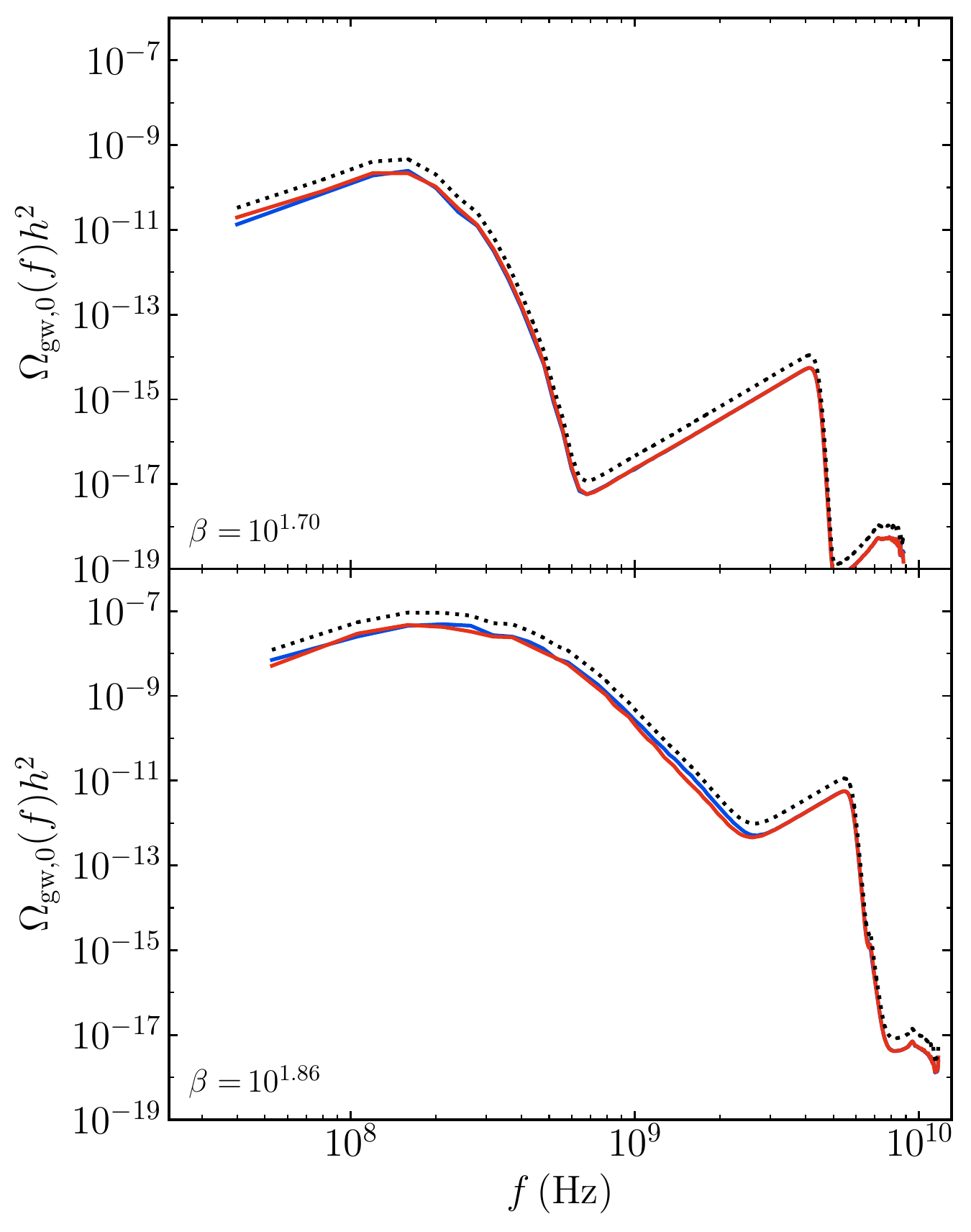}
	\caption{
		The energy density in gravitational wave polarizations today produced by an dilatonically coupled gauge field with couplings $\beta = 10^{1.7}$ (top) and $\beta = 10^{1.86}$ (bottom).
		Each frame plots the minus (red) and plus (blue) polarization and their sum (dotted black).
		}\label{fig:dilatonGWPolarization}
\end{figure}

As a first check, using the definitions in \cref{GW-pol-definition} we verify that the dilatonic coupling does not produce any appreciable polarization.
In \cref{fig:dilatonGWPolarization}, we show the final gravitational wave signal for two choices of parameters.
One parameter choice corresponds to a scenario where the universe is radiation dominated at the end of preheating---preheating completely reheats the universe---while the other corresponds to a scenario where gauge preheating is not efficient enough to completely reheat the universe.

\begin{figure*}[t]
	\includegraphics[width=1.8\columnwidth]{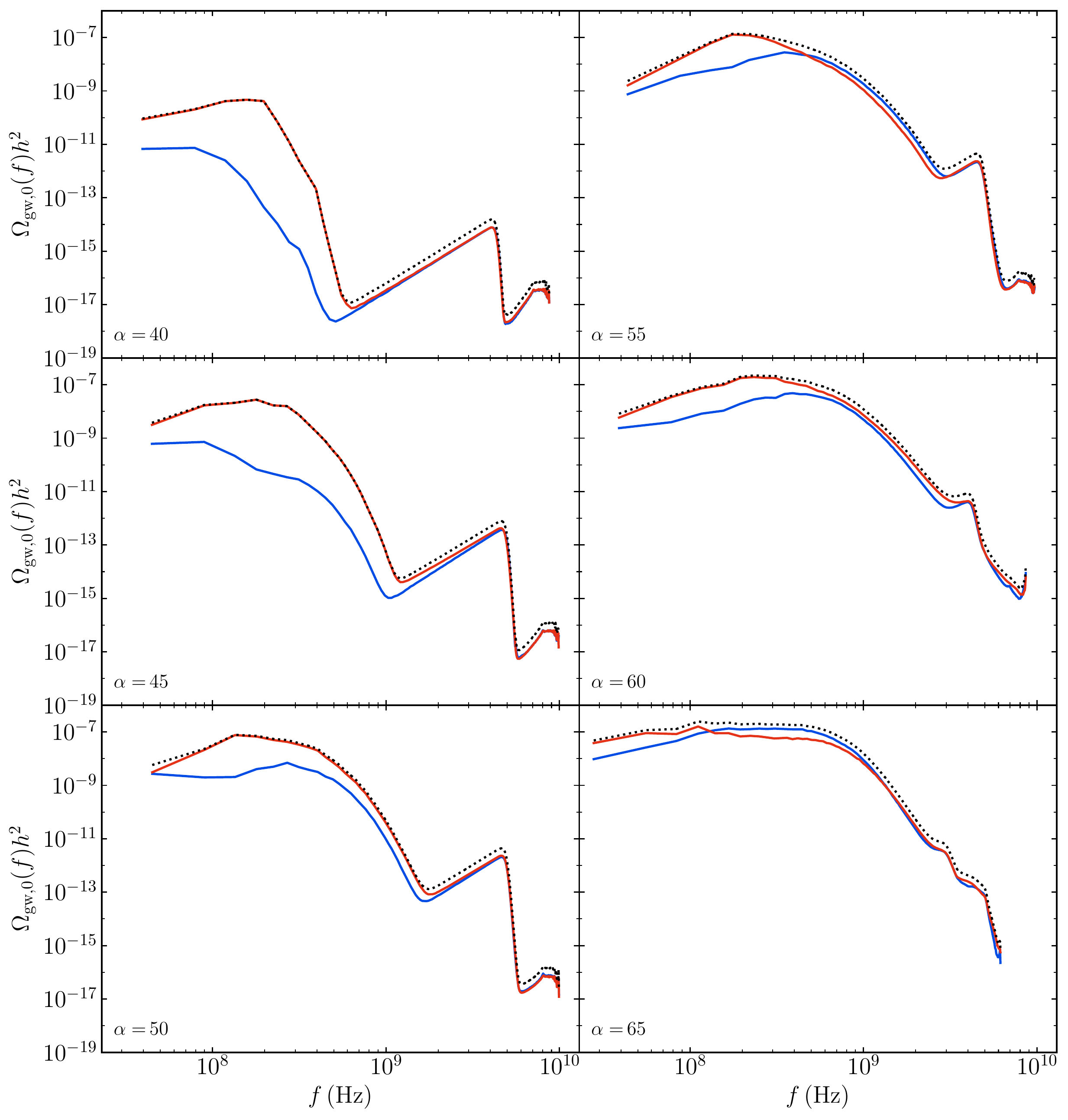}
	\caption{
		The energy density in gravitational waves today produced by an axially coupled gauge field with couplings $\alpha = 40$, $45$, $50$, $55$, $60$, and $65$ (from top to bottom, left to right; see plot labels).
		Each frame plots the minus (red) and plus (blue) polarization and their sum (dotted black).
		}\label{fig:axionGWPolarization}
\end{figure*}

In the case of the axial coupling, we see a dramatic difference.
\Cref{fig:axionGWPolarization} displays the final gravitational-wave spectra of the two polarizations and their sum for a range of couplings $\alpha$.
Indeed, on large scales the resulting gravitational wave spectrum is helically polarized as anticipated by Refs.~\cite{Sorbo:2011rz, Barnaby:2011qe}. For some simulations, rescattering of helical gauge-bosons off of the axion is strong enough that the subdominant mode is amplified significantly at smaller scales, resulting in a spectrum polarized opposite to that at large scales.

\begin{figure*}[t]
	\includegraphics[width=1.8\columnwidth]{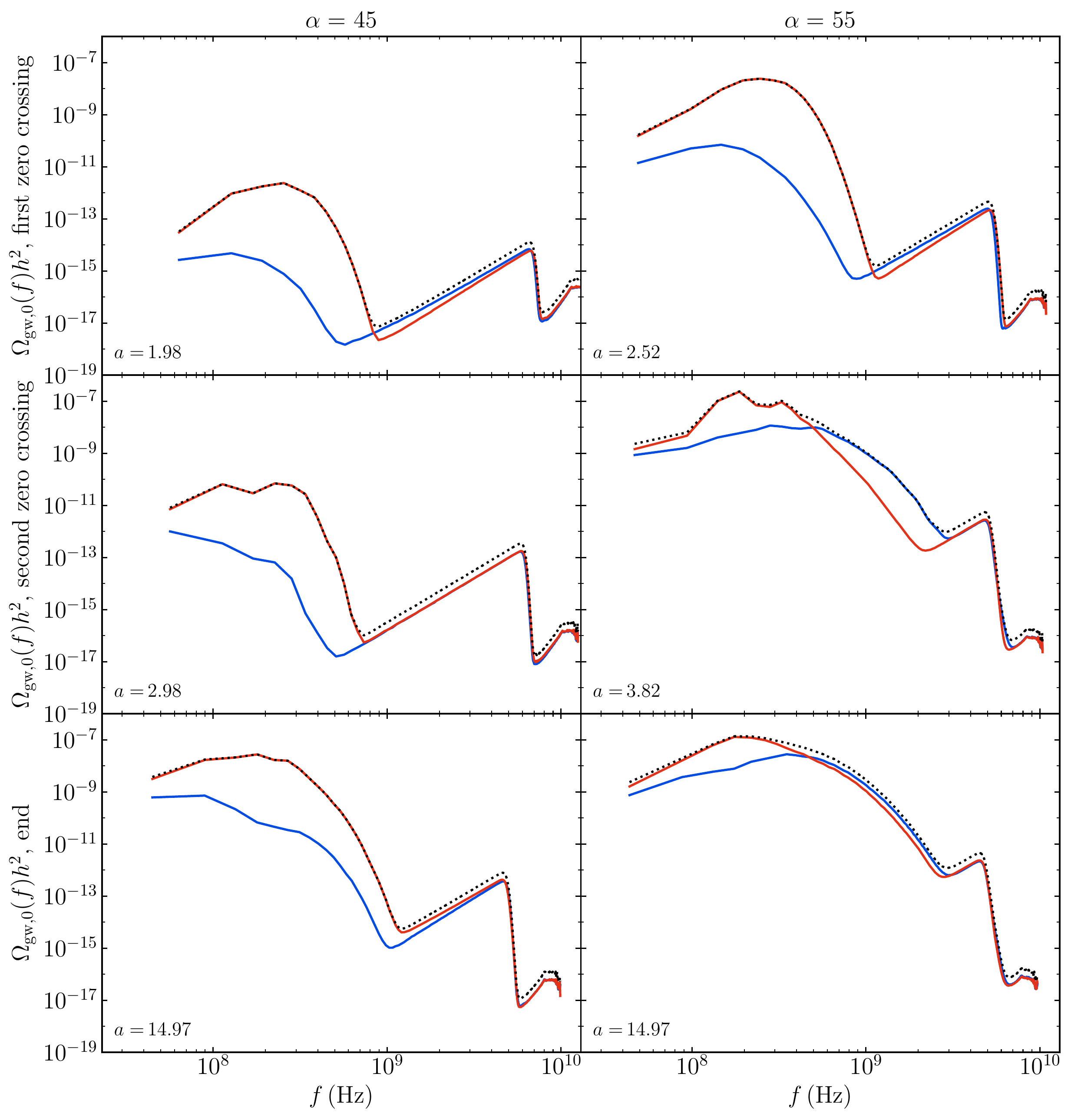}
	\caption{
		The energy density in gravitational wave polarizations today produced by an axially coupled gauge field with couplings $\alpha = 45$ (left) and $\alpha = 60$ (right).
		The top panels correspond to the first time $\phi$ crosses zero, the middle to the second zero-crossing, and the bottom panels to the end of the simulation ($a = 15$).
		Each frame plots the minus (red) and plus (blue) polarization and their sum (dotted black).
		}\label{fig:axionGWPolarizationZeroCrossings}
\end{figure*}

For a better understanding of the dynamics that lead to different final polarizations, in \cref{fig:axionGWPolarizationZeroCrossings} we plot (for two values of the coupling) the gravitational-wave spectra at the first two instances when the inflaton changes sign as well as the final spectra.
As described above, a rolling axion preferentially amplifies one helicity of the gauge field.
These oscillations lead to the alternate amplification of each gauge-boson helicity depending on the sign of the velocity of the homogeneous mode of $\phi$.
The level of amplification is exponential in the axion velocity, and the different snapshots of the spectra capture the moment of peak amplification of each gauge-boson helicity as the axion crosses the minima of its potential.
As exhibited in the top panels of \cref{fig:axionGWPolarizationZeroCrossings}, this leads to the preferential amplification of one gravitational-wave helicity since $\langle\dot{\phi}\rangle < 0$ until this point.
Comparing the middle panels, we see the amplification of the other mode as the axion velocity changes sign, $\langle\dot{\phi}\rangle > 0$.
When the coupling is larger, the subdominant mode is amplified more efficiently, and particularly so at larger scales (resulting in the opposite polarization of the IR).
This is likely due to the fact that, as the stronger couplings amplify the dominant mode more significantly with a broader resonance band, the subsequent amplification of the subdominant mode may occur through scattering from higher-frequency bosons of the dominant polarization.
Finally, in the bottom panels we see (for the stronger coupling) this skewed polarization persists to the end of the simulation as preheating completes, while for the weaker coupling, the final spectrum remains polarized on all scales.

\subsection{Gravitational leptogenesis}

One possible source of the Universe's observed matter-antimatter asymmetry is through the gravitational anomaly in the standard model lepton current,
\begin{align}
	\partial_\mu \left( \sqrt{-g} J^\mu_{B-L} \right)
	= - \frac{N_{L-R}}{24} \frac{1}{16 \pi^2} R \tilde{R}.
\end{align}
A net lepton-number is generated during inflation by the production of chiral gravitational waves which source the gravitational Pontryagin density~\cite{Alexander:2004us}.
This net lepton number is subsequently reprocessed into baryons through the hot electroweak sphaleron~\cite{Kuzmin:1985mm, Harvey:1990qw}.
Since we observe that gauge preheating after axion inflation can generate a large-amplitude gravitational wave spectrum that is significantly polarized, we now look to see whether this process produces a large enough Pontryagin density, and thus a large enough net lepton number to explain the baryon asymmetry of the Universe.

Although the axially coupled gauge fields considered here generate chiral gravitational waves during the inflationary phase~\cite{Sorbo:2011rz}, Ref.~\cite{Papageorgiou:2017yup} recently showed that the resulting lepton asymmetry is too small to explain the baryon asymmetry once the CMB bound on the tensor-to-scalar ratio is enforced.
However, the study in Ref.~\cite{Papageorgiou:2017yup} did not consider the preheating phase we consider here.

\begin{figure}[t]
	\includegraphics[width=.99\columnwidth]{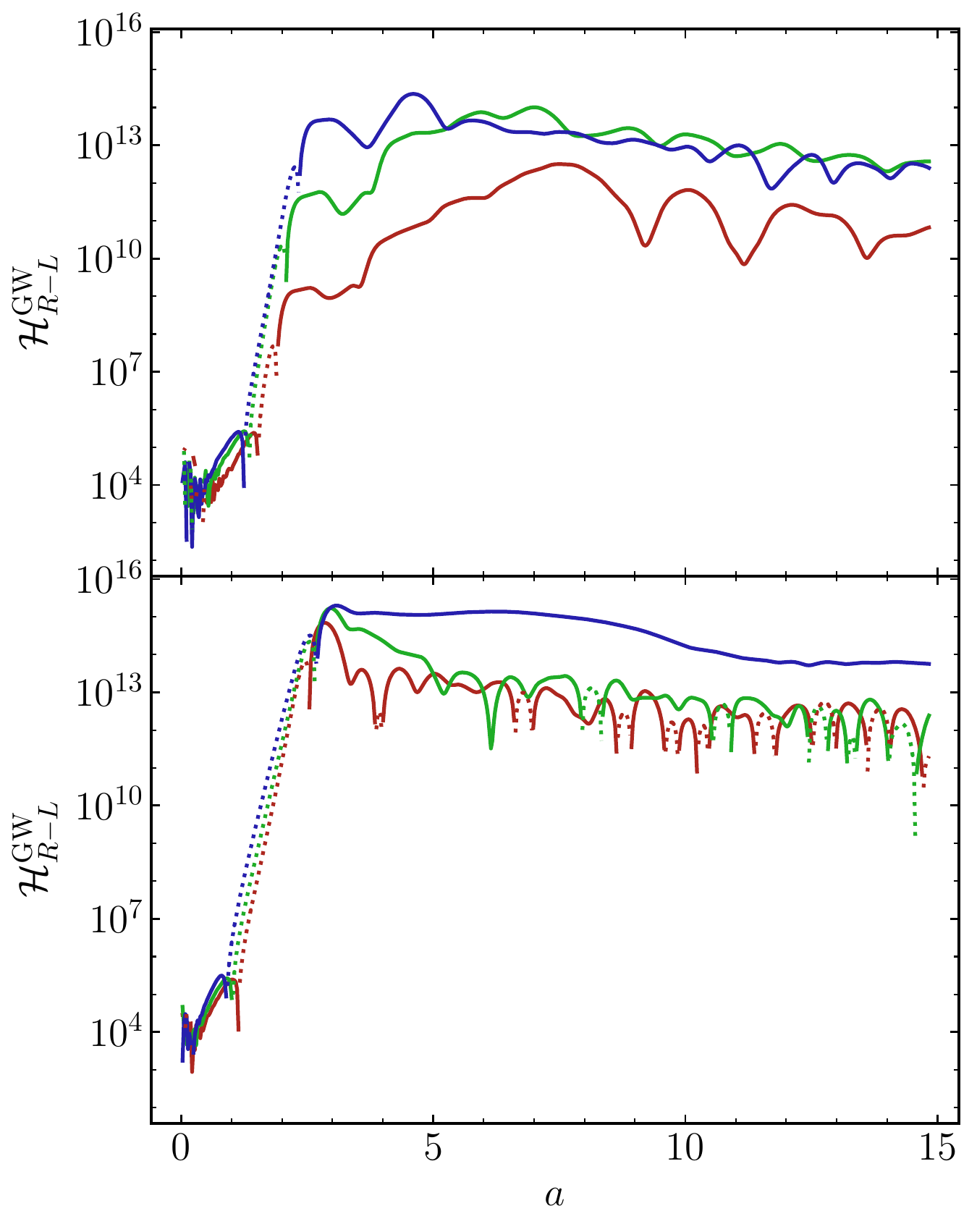}
	\caption{
		The evolution of $\mathcal{H}^\mathrm{GW}_{R-L}$ over the course of the simulations for various couplings.
		In the top panel are simulations with $\alpha = 40$ (red), $45$ (green), and $50$ (blue).
		In the bottom panel are simulations with $\alpha = 55$ (red), $60$ (green), and $65$ (blue).
		In all curves, dotted sections denote those with $\mathcal{H}^\mathrm{GW}_{R-L} < 0$.
		}\label{fig:HGWRL}
\end{figure}

To study gravitational leptogenesis during reheating, during our simulations we compute the topological charge~\cite{Adshead:2017znw}
\begin{align}
	\mathcal{H}^\mathrm{GW}_{R-L}
	\equiv \int \d \ln k \left[ \frac{k^3}{H_e^3} \frac{ \Delta_R^2 - \Delta_L^2}{H_e^2 / m_\mathrm{pl}^2} - \frac{k}{H_e} \frac{ \Delta_R'{}^2 - \Delta_L'{}^2}{H_e^4 / m_\mathrm{pl}^2} \right],
\end{align}
defining $\Delta_\lambda^2(k,\tau) \equiv (k^3 / 2 \pi^2) \left\vert h_\lambda(k,\tau) \right\vert^2$ and $\Delta_\lambda'{}^2(k,\tau) \equiv (k^3 / 2 \pi^2) \left\vert h_\lambda'(k,\tau) \right\vert^2$ and denoting the Hubble parameter at the end of inflation with $H_e$.
This integral computes the (dimensionless) expectation value of topological charge per unit Hubble volume, and leads to the net baryon-minus-lepton number per unit Hubble volume
\begin{align}
	\mathcal{N}_{B-L}(t)
	= - \frac{1}{64 \pi^2} \left( \frac{H_e}{m_\mathrm{pl}} \right)^2 \left( \mathcal{H}^\mathrm{GW}_{R-L}(t) - \mathcal{H}^\mathrm{GW}_{R-L}(t_i) \right).
\end{align}
Generating a baryon asymmetry of the right order of magnitude requires a large topological charge $\mathcal{H}^\mathrm{GW}_{R-L} \sim 10^{14}$~\cite{Adshead:2017znw}. We plot this variable for several values of the coupling $\alpha$ in \cref{fig:HGWRL}. 

These results demonstrate that the topological charge per Hubble volume generated during our simulations indeed reaches the correct order of magnitude required to achieve appreciable lepton asymmetry ($\mathcal{H}^\mathrm{GW}_{R-L} \sim 10^{14}$). However, note that the topological charge, and therefore the net lepton number, evolves during the simulation.
Accurately determining the net baryon number produced in this scenario requires solving the kinetic transport equations of the standard model of particle physics, as well as a detailed model of the neutrino mass sector~\cite{Adshead:2017znw}.
We leave this study to future work.

\section{Conclusions}\label{conclusions}

In this work we have computed the spectrum of gravitational waves produced during gauge preheating following inflation.
Dilatonic and axial couplings between a scalar (or pseudoscalar) inflaton and Abelian gauge fields produce significant gravitational radiation for coupling strengths that nearly or completely reheat the Universe.
The signals produced are remarkably loud, strong enough to be detectable by interferometers like LIGO, given an inflationary energy scale which produces gravitational waves of appropriate frequency.
Next-generation CMB experiments will be sensitive enough to $N_\mathrm{eff}$ to provide significant constraints on the couplings between the inflaton and gauge sectors.
In particular, the limit $\alpha \lesssim 50 - 55$ would be by far the strongest constraint on the axion-gauge coupling to date; prior constraints from primordial black hole production limit $\alpha \lesssim 110 - 125$ (for the inflaton potential considered here)~\cite{Linde:2012bt,Bugaev:2013fya}.

In the dilatonic model, the resulting gravitational wave spectra are unpolarized, with no preferred handedness for the resulting spectrum.
In the axion model, the production of helical gauge bosons results in a similarly polarized, parity-violating spectrum of gravitational waves, which could in principle be observed by a network of detectors~\cite{Smith:2016jqs}. Furthermore, in the axion model the chiral gravitational waves induce a topological charge that is large enough to potentially explain the baryon asymmetry of the Universe via gravitational leptogenesis, i.e., through the gravitational anomaly in the standard model lepton current. However, we find that the topological charge is not constant, even after preheating has completed. A detailed study that includes the standard model kinetic transport is required to accurately track the baryon asymmetry.

\acknowledgments

We thank Tristan Smith for useful discussions.
The work of P.A. was supported in part by a NASA Astrophysics Theory Grant NNX17AG48G.
J.T.G. is supported by the National Science Foundation Grant No. PHY-1719652.
Z.J.W. is supported in part by the United States Department of Energy Computational Science Graduate Fellowship, provided under Grant No. DE-FG02-97ER25308.
This work was performed in part at Aspen Center for Physics, which is supported by National Science Foundation Grant No. PHY-1607611.
We acknowledge NASA, the National Science Foundation, and the Kenyon College Department of Physics for providing the hardware used to carry out these simulations. 

\bibliography{GFGWs}

\end{document}